\documentclass[aps,pra,showpacs,groupedaddress,floatfix,nofootinbib]{revtex4}

\begin{document}

\title{Algebraic treatment of $\mathcal{PT}$-symmetric coupled oscillators}
\author{Francisco M. Fern\'andez}

\begin{abstract}
The purpose of this paper is the discussion of a pair of coupled linear
oscillators that has recently been proposed as a model of a system of two
optical resonators. By means of an algebraic approach we show that the
frequencies of the classical and quantum-mechanical interpretations of the
optical phenomenon are exactly the same. Consequently, if the classical
frequencies are real, then the quantum-mechanical eigenvalues are also real.
\end{abstract}


\maketitle

\affiliation{INIFTA (UNLP,CCT La Plata-CONICET), Blvd. 113 y 64 S/N, \newline
Sucursal 4, Casilla de Correo 16, 1900 La Plata, Argentina}

\section{Introduction}

\label{sec:intro}

In a recent paper Bender et al\cite{BGOPY13} discussed the classical and
quantum-mechanical versions of a system of two coupled linear oscillators
one with gain and the other with loss. When the gain and loss parameters are
equal the Hamiltonian derived from the classical equations of motion is $%
\mathcal{PT}$-symmetric and exhibits two
$\mathcal{PT}$-transitions in terms of the coupling parameter. In
the unbroken-$\mathcal{PT}$ region the classical frequencies are
real. This analysis is straightforward because one can derive
analytical expressions for such frequencies from the equations of
motion. On the other hand, the analysis of the quantum-mechanical
model does not appear to be so simple. Although the authors
obtained analytical expressions for the eigenvalues and
eigenfunctions, they estimated the regions of broken- and
unbroken-$\mathcal{PT}$ symmetry numerically and conjectured that
both the eigenvalues of the quantum-mechanical Hamiltonian and the
classical frequencies are real in the same region of model
parameters. This theoretical investigation was motivated by recent
experiments on a $\mathcal{PT}$-symmetric system of two coupled
optical resonators\cite{POLMGLFNBY13}.

The purpose of the present paper is to derive an analytic clear
connection between the classical frequencies on the one hand and
the quantum-mechanical energies on the other. In
section~\ref{sec:model} we apply a well known algebraic method and
show that the quantum-mechanical frequencies (spacing between
eigenvalues) are exactly the classical ones. We also write the
quantum-mechanical energies in terms of the classical frequencies
and analyse the spectrum. Finally, in
section~\ref{sec:conclusions} we review the main results of the
paper and draw conclusions.

\section{Quantum-mechanical model}

\label{sec:model}

As indicated above, the analysis of the classical model is straightforward
and here we focus on the quantal version in the case of equal gain and loss.
The Hamiltonian operator derived from the classical equations of motion is
given by\cite{BGOPY13}
\begin{equation}
H=p_{x}p_{y}+\gamma (yp_{y}-xp_{x})+\left( \omega ^{2}-\gamma ^{2}\right) xy+%
\frac{\epsilon }{2}\left( x^{2}+y^{2}\right) ,  \label{eq:H}
\end{equation}
where $p_{x}$ and $p_{y}$ are the momenta conjugate to coordinates $x$ and $y
$: $[x,p_{x}]=[y,p_{y}]=i$. It is $\mathcal{PT}$ symmetric $\mathcal{PT}H%
\mathcal{PT}=H$ since it is invariant under the combined effect of parity $%
\mathcal{P}:\{x,y,p_{x},p_{y}\}\rightarrow \{-y,-x,-p_{y},-p_{x}\}$ and time
reversal $\mathcal{T}:\{x,y,p_{x},p_{y}\}\rightarrow \{x,y,-p_{x},-p_{y}\}$%
\cite{BGOPY13}. Note that we can also choose $\mathcal{P}:\{x,y,p_{x},p_{y}%
\}\rightarrow \{y,x,p_{y},p_{x}\}$ for exactly the same purpose. In the
language of point-group symmetry such \textit{parity} transformations are
given by reflection planes $\sigma _{v}$ and $\sigma _{v}^{\prime }$
perpendicular to the $x-y$ plane\cite{C90,T64}. Since the Hamiltonian $H$ is
also invariant under inversion $\hat{\imath}:\{x,y,p_{x},p_{y}\}=%
\{-x,-y,-p_{x},-p_{y}\}$ its eigenvectors $\left| \psi \right\rangle $
satisfy $\hat{\imath}\left| \psi \right\rangle =\pm \left| \psi
\right\rangle $. This result is consistent with the solutions of the form $%
P_{mn}(x,y)e^{-(2axy+bx^{2}+cy^{2})}$ obtained by Bender et
al\cite{BGOPY13} in the coordinate representation, where the
polynomials $P_{mn}(x,y)$ satisfy
$P_{mn}(-x,-y)=(-1)^{m}P_{mn}(x,y)$. Point-group symmetry proved
to be useful for the discussion of broken- and
unbroken-$\mathcal{PT}$ symmetry in some anharmonic
oscillators\cite{KC08,FG14}. As far as we know Klaiman and
Cederbaum\cite{KC08} were the first to construct a non-Hermitian
Hamiltonian with consecutive $\mathcal{PT}$ phase transitions in
terms of the coupling parameter.

It is worth noting that the Hamiltonian (\ref{eq:H}) is also a
symmetric operator $\left\langle H\psi \right. \left| \varphi
\right\rangle =\left\langle \psi \right. \left| H\varphi
\right\rangle $, which for brevity we formally express by means of
the usual notation for Hermitian operators as $H^{\dagger }=H$. In
particular, note that $(yp_{y}-xp_{x})^{\dagger
}=(p_{y}y-p_{x}x)=(yp_{y}-xp_{x})$. Therefore, one expects real
eigenvalues at least for some range of the model parameters
$\omega $, $\gamma $ and $\epsilon $.

In order to solve the eigenvalue equation $H\left| \psi \right\rangle
=E\left| \psi \right\rangle $ in a way that clearly reveals the connection
with the classical interpretation we resort to a well known algebraic method%
\cite{FC96}. It is suitable when there exists a set of symmetric operators $%
\{O_{1},O_{2},\ldots ,O_{N}\}$ that satisfy the commutation relations
\begin{equation}
\lbrack H,O_{i}]=\sum_{j=1}^{N}H_{ji}O_{j}.  \label{eq:[H,Oi]}
\end{equation}
We look for an operator of the form
\begin{equation}
Z=\sum_{i=1}^{N}c_{i}O_{i},  \label{eq:Z}
\end{equation}
such that
\begin{equation}
\lbrack H,Z]=\lambda Z.  \label{eq:[H,Z]}
\end{equation}
The operator $Z$ is important for our purposes because
\begin{equation}
HZ\left| \psi \right\rangle =(E+\lambda )Z\left| \psi \right\rangle .
\label{eq:HZ|Psi>}
\end{equation}

It follows from equations (\ref{eq:[H,Oi]}), (\ref{eq:Z}) and (\ref{eq:[H,Z]}%
) that
\begin{equation}
(\mathbf{H}-\lambda \mathbf{I})\mathbf{C}=0,  \label{eq:HC=-lambdaC}
\end{equation}
where $\mathbf{H}$ is an $N\times N$ matrix with elements $H_{ij}$, $\mathbf{%
I}$ is the $N\times N$ identity matrix, and $\mathbf{C}$ is an $N\times 1$
column matrix with elements $c_{i}$. $\mathbf{H}$ is called the adjoint or
regular matrix representation of $H$ in the operator basis $%
\{O_{1},O_{2},\ldots ,O_{N}\}$\cite{FC96}. In the case of an Hermitian
operator we expect all the roots $\lambda _{i}$, $i=1,2,\ldots ,N$ to be
real. These roots are obviously the natural frequencies of the
quantum-mechanical system. Here we apply the same approach to symmetric
Hamiltonians because all the relevant equations are formally identical. If $%
\lambda $ is real then it follows from equation (\ref{eq:[H,Z]}) that
\begin{equation}
\lbrack H,Z^{\dagger }]=-\lambda Z^{\dagger },  \label{eq:[H,Z+]}
\end{equation}
where $Z^{\dagger }$ is a linear combination like (\ref{eq:Z})
with coefficients $c_{i}^{*}$. This equation tells us that if
$\lambda $ is a real root of $\det (\mathbf{H}-\lambda
\mathbf{I})=0$, then $-\lambda $ is also a root. Obviously, $Z$
and $Z^{\dagger }$ are a pair of annihilation-creation or ladder
operators because, in addition to (\ref{eq:HZ|Psi>}), we also have
\begin{equation}
HZ^{\dagger }\left| \psi \right\rangle =(E-\lambda )Z^{\dagger }\left| \psi
\right\rangle .  \label{eq:HZ+|Psi>}
\end{equation}
Other authors have already applied Lie-algebraic methods to the Hamiltonian (%
\ref{eq:H}) without coupling ($\epsilon =0$)\cite{D81, CRV92, CJ06} but they
were not interested in the quantum-mechanical frequencies.

In the present case, the obvious choice $\{O_{1},O_{2},O_{3},O_{4}\}=%
\{x,y,p_{x},p_{y}\}$ leads to the matrix representation
\begin{equation}
\mathbf{H}=i\left(
\begin{array}{llll}
\gamma & 0 & \epsilon & \omega ^{2}-\gamma ^{2} \\
0 & -\gamma & \omega ^{2}-\gamma ^{2} & \epsilon \\
0 & -1 & -\gamma & 0 \\
-1 & 0 & 0 & \gamma
\end{array}
\right) ,
\end{equation}
with characteristic polynomial

\begin{equation}
\lambda ^{4}+\lambda ^{2}\left( 4\gamma ^{2}-2\omega ^{2}\right) -\epsilon
^{2}+\omega ^{4}=0,  \label{eq:f(lambda)=0}
\end{equation}
that is exactly the one that yields the classical frequencies\cite{BGOPY13}.
Two of its roots are
\begin{eqnarray}
\lambda _{1} &=&\sqrt{\sqrt{\epsilon ^{2}+4\gamma ^{4}-4\gamma ^{2}\omega
^{2}}-2\gamma ^{2}+\omega ^{2}}  \nonumber \\
\lambda _{2} &=&\sqrt{-\sqrt{\epsilon ^{2}+4\gamma ^{4}-4\gamma ^{2}\omega
^{2}}-2\gamma ^{2}+\omega ^{2},}  \label{eq:lambda1,lambda2}
\end{eqnarray}
and the other two ones are $\lambda _{3}=-\lambda _{1}$ and $\lambda
_{4}=-\lambda _{2}$ in agreement with the more general equations (\ref
{eq:[H,Z]}) and (\ref{eq:[H,Z+]}). The operators $Z_{1}$ and $Z_{2}$
associated to $\lambda _{1}$ and $\lambda _{2}$ are creation or rising,
while $Z_{2}=Z_{1}^{\dagger }$ and $Z_{3}=Z_{2}^{\dagger }$ are annihilation
or lowering. The classical and quantal frequencies are exactly the same
because the relevant Poisson brackets and commutators are similar: $%
i\{H,O_{i}\}\rightarrow [H,O_{i}]$. This result reveals why the condition
for real classical frequencies
\begin{equation}
2\gamma \sqrt{\omega ^{2}-\gamma ^{2}}<\epsilon <\omega ^{2}
\label{eq:conditions}
\end{equation}
is also the condition for real spectrum (unbroken-$\mathcal{PT}$ region) in
the quantum-mechanical counterpart\cite{BGOPY13}. If we write the polynomial
equation (\ref{eq:f(lambda)=0}) as $\left( \lambda ^{2}-\lambda
_{1}^{2}\right) \left( \lambda ^{2}-\lambda _{2}^{2}\right) =0$ then we
realize that
\begin{eqnarray}
\lambda _{1}^{2}+\lambda _{2}^{2} &=&2\omega ^{2}-4\gamma ^{2}  \nonumber \\
\lambda _{1}^{2}\lambda _{2}^{2} &=&\omega ^{4}-\epsilon ^{2}.
\label{eq:lam1^2lam2^2}
\end{eqnarray}

Throughout this paper we keep the parameter $\omega $ in order to facilitate
the discussion of the results of Bender et al. However, it is worth noting
that we can choose $\omega =1$ without loss of generality as follows from
the transformation $\{\lambda ,a,\omega ,\gamma ,\epsilon \}\rightarrow \{%
\frac{\lambda }{\omega },\frac{a}{\omega },1,\frac{\gamma }{\omega },\frac{%
\epsilon }{\omega ^{2}}\}$.

Bender et al\cite{BGOPY13} derived the energies
\begin{equation}
E_{mn}=(m+1)a+(2n-m)\Delta ,  \label{eq:Emn}
\end{equation}
where $m=0,1,\ldots $ and $n=0,1,\ldots ,m$. In this equation $a$ is a root
of

\begin{equation}
4a^{4}+4a^{2}\left( 2\gamma ^{2}-\omega ^{2}\right) +\epsilon ^{2}+4\gamma
^{2}\left( \gamma ^{2}-\omega ^{2}\right) =0,  \label{eq:f(a)=0}
\end{equation}
and
\begin{eqnarray}
\Delta &=&\sqrt{bc-\gamma ^{2}},  \nonumber \\
b &=&c^{*}=\frac{\epsilon }{2(a+i\gamma )}.  \label{eq:Delta,b,c}
\end{eqnarray}
The expressions for $a$, $b$ and $c$ come from solving the eigenvalue
equation $H\left| \psi _{00}\right\rangle =E_{00}\left| \psi
_{00}\right\rangle $ in the coordinate representation with the ansatz $\psi
_{00}(x,y)=e^{-(bx^{2}+cy^{2}+2axy)}$, procedure that also yields $E_{00}=a$%
. If we write $\xi =a^{2}$ then we obtain the roots
\begin{eqnarray}
\xi _{1} &=&\frac{\omega ^{2}-2\gamma ^{2}-\sqrt{\omega ^{4}-\epsilon ^{2}}}{%
2},  \nonumber \\
\xi _{2} &=&\frac{\omega ^{2}-2\gamma ^{2}+\sqrt{\omega ^{4}-\epsilon ^{2}}}{%
2}.  \label{eq:xi1,xi2}
\end{eqnarray}
Following Bender et al we write $a_{1}=-\sqrt{\xi _{1}}$,
$a_{2}=\sqrt{\xi _{1}}$, $a_{3}=-\sqrt{\xi _{2}}$,
$a_{4}=\sqrt{\xi _{2}}$. For concreteness, from now on we choose
\begin{equation}
a=a_{2}=\frac{1}{2}\sqrt{2\omega ^{2}-4\gamma ^{2}-2\sqrt{\omega
^{4}-\epsilon ^{2}}}.
\end{equation}

So far, we have shown that the classical and quantum-mechanical
interpretations exhibit exactly the same frequencies; it only remains to
rewrite the eigenvalues (\ref{eq:Emn}) in terms of these frequencies. One
can easily verify that
\begin{equation}
\lambda _{1}=\Delta +a,\,\lambda _{2}=\Delta -a,  \label{eq:lambda(Delta,a)}
\end{equation}
is consistent with (\ref{eq:lam1^2lam2^2}). Thus, the expression for the
energies becomes

\begin{eqnarray}
E_{m,n} &=&\frac{\lambda _{1}\left( 2n+1\right) }{2}-\frac{\lambda
_{2}\left( 2m-2n+1\right) }{2}=  \nonumber \\
&=&n\lambda _{1}+(n-m)\lambda _{2}+a.  \label{eq:Emn(lambda1,lambda2)}
\end{eqnarray}

Bender et al\cite{BGOPY13} showed numerically that $a_{2}$ and $\Delta $ are
real and positive in the unbroken-$\mathcal{PT}$ region and concluded that
the eigenvalues are also real and positive. However, this is not the case
because $a-\Delta <0$ and for every value of $n$ $E_{mn}\rightarrow -\infty $
as $m\rightarrow \infty $. As an example, consider the model parameters $%
\omega =1$, $\gamma =0.05$ and $\epsilon =0.5$ that lie the unbroken-$%
\mathcal{PT}$ region. In this case $\Delta \approx 0.964630863$ and $%
a\approx 0.2539434939$ that confirms what we have just said.

We can analyse those results by means of the algebraic method. If $E_{00}$
were the lowest eigenvalue then both $Z_{3}\psi _{00}$ and $Z_{4}\psi _{00}$
would be expected to vanish. However, for the parameters chosen above we
found that $Z_{2}\psi _{00}$ and $Z_{3}\psi _{00}$ vanish while $Z_{1}\psi
_{00}$ and $Z_{4}\psi _{00}$ do not. Therefore, the eigenfunctions are given
by
\begin{equation}
\psi _{nk}=Z_{1}^{n}Z_{4}^{k}\psi _{00}  \label{eq:psi_n,k}
\end{equation}
with eigenvalues $E_{nk}=n\lambda _{1}-k\lambda _{2}+a$ wich agree with (\ref
{eq:Emn(lambda1,lambda2)}) if $k=m-n$. The algebraic method clearly shows
that the spectrum is unbounded from below.

\section{Conclusions}

\label{sec:conclusions}

The main purpose of this paper is to show that part of the
mathematical analysis of some classical systems also applies to
their quantum-mechanical counterparts. The underlying connection
is that the frequencies of both interpretations are exactly the
same because of the similarity between the Poisson brackets and
commutators. Therefore, if the frequencies of the motion of the
classical system are real then the quantum-mechanical eigenvalues
are also real. The quantum-mechanical frequencies are the
eigenvalues of the regular or adjoint matrix representation of the
Hamiltonian operator in a suitable basis set of operators, whereas
the corresponding eigenvectors provide the ladder operators. This
well known algebraic approach is suitable for many
problems\cite{FC96} and in particular for Hamiltonians that are
quadratic functions of the coordinates and their conjugate
momenta. Such Hamiltonians are suitable models for many physical
problems like the one that motivated the paper by Bender et
al\cite{BGOPY13}, among others\cite{D81, CRV92, CJ06}.

In closing it is worth mentioning that the fact that the spectrum
of the Hamiltonian (\ref{eq:H}) is not strictly positive, contrary
to what Bender et al assumed, does not appear to be relevant to
the interpretation of the physical data which is fitted by the
classical (and also quantal) frequencies\cite{POLMGLFNBY13}.

\end{document}